\documentstyle[prl,aps,epsfig,twocolumn]{revtex}

\draft

\newlength{\textwidthm}
\begin{document}
\title{Coherent Atom Interactions Mediated by Dark-State Polaritons}
\author{A. Andr\'{e}$^1$, L.-M. Duan$^2$ and M.D. Lukin$^1$}
\address{$^1$ITAMP, Harvard-Smithsonian Center for Astrophysics and \\
Physics Department, Harvard University, Cambridge, MA~~02138 \\
$^2$Institute for Theoretical Physics, University of Innsbruck, Austria.}
\date{\today}
\maketitle
\begin{abstract}
We suggest a technique to induce effective, controllable 
interactions between atoms that is based on Raman scattering 
into an optical mode propagating with a slow group velocity. The resulting 
excitation corresponds to the creation of spin-flipped atomic pairs
in a way that is analogous to correlated photon emission in 
optical parametric amplification.  The technique can be used for 
fast generation of entangled  atomic ensembles, spin squeezing and 
applications in quantum information processing. 
\end{abstract}



The intriguing possibility for controlled manipulation of 
interacting quantum systems  is the basis for a number of exciting 
developments in the field of quantum information science \cite{Q-info}. 
These are 
expected to have an impact in a broad area ranging from 
quantum computation and quantum communication \cite{Q-comm} to  
precision measurements \cite{Prec-meas} and controlled modeling of complex 
quantum phenomena \cite{complex}.

This Letter describes a new technique to induce effective coherent 
interactions between atoms in metastable states. 
The technique is based on a resonantly enhanced
nonlinear process involving Raman scattering into a 
``slow'' optical mode \cite{slow-light}, which creates a 
pair of spin-flipped atom and slowly propagating coupled excitation
of light and matter (dark-state  polariton). When the group
velocity of the polariton is reduced to zero \cite{darkpolar,stop-light}, 
this results 
in pairs of spin flipped atoms.

The present phenomenon of spin pair creation 
exhibits strong similarities with optical parametric 
amplification (OPA), in which pairs of photons are generated that 
possess non-classical  correlations in photon number, 
quadrature component fluctuations or in polarization states \cite{walls}. 
In direct
analogy, the present technique is capable of generating 
non-classically correlated atomic ensembles and 
entangled  spin excitations. The latter can easily be converted 
into  corresponding states of photon wavepackets ``on demand'', 
which makes  the present approach most suitable for implementing protocols 
in quantum information  processing that require a combination of 
deterministic sources of entangled states and 
long-lived quantum memory \cite{Q-mem,Qcomm-Raman}.   

The present technique can also be viewed as a new mechanism for 
coherent ``collisions'' \cite{meystre}
between atoms mediated by light. In particular, the case when 
atomic pairs are excited into two different levels 
(as e.g. in Fig.1a) closely resembles coherent spin-changing 
interactions that occur in degenerate atomic samples \cite{degen-atom}, 
whereas the case
when atomic pairs are stimulated into identical states (Fig.1b) 
is reminiscent of dissociation of a molecular condensate \cite{molcond}. 
To put this analogy in perspective we note 
that the rate of the present optically induced process can exceed that of 
weak interatomic interactions by orders of magnitude. Therefore the 
present work may open up interesting new possibilities 
for studying  many-body phenomena of strongly interacting atoms. 

Before proceeding we note that a 
number of proposals 
have been made for generating entangled states of atomic ensembles 
and resulting in so-called spin 
squeezed states.  Some are based 
on interatomic interactions at ultra-cold temperatures \cite{squ-cold}, whereas
others involve  mapping the states of non-classical light fields into 
atoms \cite{Polzik}, QND measurements of spins \cite{qnd} with light and 
Rydberg blockade \cite{Bouchoule}. 
In contrast to these mechanisms 
the present approach does not require  
coherence of the atomic motion or sources of non-classical light 
and  is completely deterministic thereby 
significantly simplifying possible experimental realizations. 
We further show that the present technique can be made robust with 
respect to realistic decoherence processes such as spontaneous emission and 
leakage of slow photons from the medium.
We also note the work of Franson and co-workers \cite{Franson} 
on quantum logic based on 
so-called ``photon-exchange'' interaction. However further analysis 
\cite{Franson2} showed that
these mechanisms do not result in a non-linear interaction.

We consider a system of $N$ atoms (Fig.1) interacting with two 
classical driving fields and
one quantized mode that is initially in a vacuum state. 
Relevant atomic sublevels
include two manifolds of metastable states (e.g hyperfine sublevels of 
electronic ground state) and excited states that might be accessed by optical 
transitions. The atoms are initially prepared in their ground states 
$|g\rangle$. One of the classical fields (Rabi frequency $\Omega_1$) 
is detuned from the atomic resonance by an amount roughly equal to the 
frequency splitting between ground state manifolds. The other (Rabi frequency $\Omega_2$) is resonant with an atomic transition 
$|b_2\rangle\rightarrow|a_2\rangle$. 
The quantized field can be involved in two Raman transitions 
corresponding to   Stokes and anti-Stokes processes. Whereas the former
corresponds to the usual Stokes scattering in the forward direction, the 
latter 
establishes an Electromagnetically Induced Transparency (EIT) and slows
down its group velocity.
The pair excitation can be viewed as resulting from quantized photon 
exchange between atoms (Fig.2) in a two-step process.  The first flipped
spin is created due to  Stokes Raman scattering, which also results in 
photon emission in a corresponding Stokes mode. In the presence of EIT,
this
photon is directly converted into a dark-state polariton which becomes 
purely atomic  when the group velocity is  reduced to zero.  
This implies that atomic spins are always flipped in pairs. 
In Fig.1a the two final states involved in Raman 
transitions are different and atomic pairs in different states
are created. In Fig.1b the final states of the two Raman processes are 
identical, in which case  atomic pairs in the same state result.

In what follows we will focus on a system (Fig.1a) involving two 
atomic modes. Consideration of the scheme of Fig.1b proceeds along 
the same lines.  
For conceptual simplicity we here assume that the quantized field 
corresponds to  a single mode  of a running-wave cavity with a creation 
operator $\hat{a}^\dagger$ and atom-field coupling constants $g_1$ and $g_2$. 
Generalization to multi-mode i.e. travelling wave configuration is 
straightforward. 
The interaction
Hamiltonian for the system of N atoms and light can be split into two 
parts $H = H_{ram} + H_{res}$, which are given by: 
\begin{eqnarray}
H_{ram}=&-&\hbar\Delta \Sigma_{a1a1} - \hbar\delta_1 \Sigma_{b_1b_1} 
\nonumber \\
&+&[\hbar\Omega_1 \Sigma_{ga1} 
+ \hbar g_1 a^\dagger \Sigma_{b_1a1} + {\rm
h.c.}], \label{ham1} \\
H_{res}=&&\hbar\delta_2 \Sigma_{b_2b_2} + \hbar\delta_2 \Sigma_{a_2a_2} 
\nonumber \\
&+& [\hbar g_2 a^\dagger \Sigma_{ga2} + \hbar\Omega_2 \Sigma_{b_2a2} + {\rm h.c.}], \label{ham2}
\end{eqnarray}
where $\Sigma_{\mu\nu} = \sum_i 
|\mu\rangle_{ii}\langle \nu|$ are collective atomic operators corresponding to 
transitions between atomic states $|\mu\rangle,|\nu\rangle$, $\Delta$ is the
detuning of the classical field $\Omega_1$ from the single-photon transition 
$|g\rangle \rightarrow |a_1\rangle$, $\delta_1$ and $\delta_2$ are the 
two-photon detunings from the $|g\rangle \rightarrow |b_1\rangle$ and
$|g\rangle \rightarrow |b_2\rangle$ transitions respectively as shown in
Fig.1.

In the limit of large detuning $\Delta$ and ignoring two-photon detunings for 
the moment, the Hamiltonian $H_{ram}$ describes
an off-resonant Raman scattering. After a canonical transformation 
corresponding to adiabatic elimination of the excited 
state $H_{ram}$ becomes equivalent to: 
\begin{eqnarray}
H_{ram}= \hbar \sqrt{N} { g_1^* \Omega_1 \over \Delta}  a S_1 
+ {\rm h.c.} \label{ham3}, 
\end{eqnarray}
where we disregarded the light shift $\delta_L=\Omega_1^2/\Delta$ and 
introduced  
$S_1 = 1/\sqrt{N} \Sigma_{gb_1}$. The light shift $\delta_L$ can be easily compensated by re-defining the energy of the atomic levels and will be disregarded in the remainder of this Letter.  The resonant part of 
the Hamiltonian $H_{res}$ is best analyzed in terms of dark and bright-state 
polaritons
\begin{eqnarray}
P_D &=& {\Omega_2 a - g_2 \sqrt{N} S_2 \over \sqrt {g_2^2 N + \Omega_2^2}}, \nonumber \\
P_B &=& {g_2 \sqrt{N} a + \Omega_2 S_2 \over \sqrt {g_2^2 N + \Omega_2^2}},  
\label{polaritons}
\end{eqnarray}
which are superpositions of photonic and atomic excitations, with 
$S_2 = 1/\sqrt{N} \Sigma_{gb_2}$. In particular, $H_{res}$
has an important family of dark-states: 
\begin{eqnarray}
|D^n\rangle \sim (P_D^\dagger)^n |g\rangle|{\rm vac}\rangle
\label{dark}
\end{eqnarray}
with zero eigenenergies. Note that all other eigenstates of $H_{res}$ 
have, in general,  non-vanishing interaction energy. Under conditions of 
Raman resonance and sufficiently slow excitation (``adiabatic condition'') 
the Stokes photons emitted by Raman scattering, Eq.(\ref{ham3}),  will 
therefore couple  solely to the dark-states (\ref{dark}). In this case
the evolution of the entire system is described by an effective 
Hamiltonian: 
\begin{eqnarray}
H_{eff} = \hbar \xi (P_D S_1 + S_1^\dagger P_D^\dagger), 
\label{eff} 
\end{eqnarray}
with $\xi = \Omega_1 \Omega_2/ \Delta  \times g_1^* \sqrt{N}/  
\sqrt{|g_2|^2 N + \Omega_2^2}$. 
The Hamiltonian (\ref{eff}) describes the coherent
process of generation of pairs involving polaritons and spin-flipped 
atoms. 
Note that for a small number of excitations the spin waves and
polaritons
obey bosonic commutation relations and this Hamiltonian  is formally 
equivalent to that describing optical parametric amplification (OPA) of two
modes
\cite{walls}. 
This is also analogous to the ``counter-twisting'' model of Ref.
\cite{2axis-squ}, which is known to result in maximal spin squeezing 
for large number of excitations.

We now consider the scenario in which the system is evolving for a time $\tau$ 
under the Hamiltonian $H_{eff}$, after which both fields are turned off. 
If the  procedure is adiabatic  upon turn-off of the coupling fields 
$\Omega_{1,2}$ the polaritons are converted into pure spin excitations 
$P_D \rightarrow S_2$. Hence the entire procedure will correspond to the 
following state of the system:
\begin{eqnarray}
|\Psi\rangle = && { 1\over \cosh\xi\tau} \sum_n (\tanh\xi\tau)^n {1 \over n!}
(P_D^\dagger)^n (S_1^\dagger)^n |g\rangle|{\rm vac}\rangle \nonumber \\
 && \rightarrow 
{ 1\over \cosh\xi\tau} \sum_n (\tanh\xi\tau)^n |n_{b_1},n_{b_2} \rangle|{\rm vac}\rangle.
\end{eqnarray}
Here $|n_{b_1},n_{b_2} \rangle = 1/n! (S_2^\dagger)^n (S_1^\dagger)^n |g\rangle$ are 
Dicke-like 
symmetric states of atomic ensemble and we assumed $n_{b_1,b_2} \ll N$.
For non-zero $\xi \tau$ this state describes an entangled state, 
for which relative fluctuations between the two modes 
decreases exponentially to values well below the standard quantum limit 
(SQL) corresponding to uncorrelated atoms.

The above analysis only includes the interaction with a single 
(forward-propagating) quantized radiation mode and neglects 
decoherence processes. 
We now take into account realistic decoherence
mechanisms such as spontaneous emission from the 
excited states in all directions and decay of the cavity mode 
with a rate $\kappa$. The evolution of atomic operators is then described 
by Heisenberg-Langevin equations:
\begin{eqnarray}
{\dot \Sigma}_{\mu\nu} = -\gamma_{\mu\nu} \Sigma_{\mu\nu} + {i \over \hbar}
[H, \Sigma_{\mu\nu}] + F_{\mu\nu}, 
\end{eqnarray}
where $\gamma_{\mu\nu}$ is a  decay rate of coherence $\mu\rightarrow \nu$
and $F_{\mu\nu}$ are associated noise forces. The latter have zero average 
and are $\delta$-correlated with associated diffusion coefficients that 
can be found using the Einstein relations. 
The cavity mode obeys the equation of motion:
\begin{equation}
\dot{a} = -\kappa a -i g_1 \Sigma_{b_1a_1} -i g_2\Sigma_{ga_2}+F_a(t).
\end{equation}

We proceed by adiabatic elimination of optical polarizations associated with
Stokes emission. To this end we assume large single-photon 
detuning $\Delta \gg \gamma$ and to first order in
$\hat{a}$ we obtain the following equations of motion
for the metastable coherences
\begin{eqnarray}
\dot{S_1}^\dagger &=& -\left[{\bar \gamma_{gb}}+i\delta_1\right] S_1^\dagger +
i\frac{g_1^*\sqrt{N}\Omega_1}{\Delta}\,a + {\bar F}_{S_1}^\dagger \\ \nonumber 
\dot{S}_2&=&-\left[{\bar \gamma}_{gb}+i\delta_2\right] S_2 -i(\Omega_2/\sqrt{N})\Sigma_{ga_2} + {\bar F}_{S_2}
\label{groundco}
\end{eqnarray}
where ${\bar \gamma}_{gb} = \gamma_{gb} + \gamma_L$ and $\gamma_L=\gamma_{ag}|\Omega_1|^2/\Delta^2$ is the optical pumping rate. The ground state coherence
thus decays at a rate modified by isotropic spontaneous emission from the 
excited state (we consider the case $\gamma_L\gg\gamma_{gb}$).

To treat the resonant EIT-like interaction we first rewrite the equations of 
motion in terms of the dark and bright polariton operators (\ref{polaritons})
and proceed to adiabatically eliminate the optical coherence
$\Sigma_{ga_2}$ and the bright state polariton $P_B$.
In the relevant limit when $|g_2|^2N/\gamma_{ag}\kappa \gg 1$ and when 
$\eta= |g_2|^2N/|\Omega_2|^2$ the ratio of vacuum light velocity to 
group velocity is large $\eta\gg 1$, we find:

\begin{eqnarray}
\dot{P}_D &=& -[\frac{\kappa+\eta (\gamma_L+i\delta_2)}{1+\eta}]P_D +i \xi 
S_1^\dagger + \tilde{F}_D(t), \nonumber \\
\dot{S}_1^\dagger &=& [\frac{|g_2|^2}{|g_1|^2}\gamma_L-\gamma_L-i\delta_1] 
S_1^\dagger -i \xi P_D + \tilde{F}_{S_1}^\dagger(t).
\end{eqnarray}

We note that cavity losses are strongly suppressed in the limit $\eta \gg 1$: subsequent to the large group velocity reduction \cite{slow-light}, 
the polariton is almost purely atomic and the excitation leaks very slowly 
out of the medium. 

The equation of motion for coherence $S_1^\dagger$ contains 
a loss term 
(due to isotropic spontaneous emission) and a linear gain term (due to emission
into bright polariton). 
The two can compensate each other. However the linear phase-insensitive 
amplification is also accompanied by
correspondingly increased fluctuations, represented by new Langevin forces 
$\tilde{F}_D(t),\tilde{F}_{S_1}^\dagger(t)$. Terms resulting from the direct 
coupling of dark and bright polaritons can be neglected for $|\Omega_2^2|\gg \gamma_{ag}\kappa$, which is reminiscent of the condition under which EIT is established.

To quantify the resulting quantum correlations we introduce a squeezing 
parameter in direct analogy to the optical parametric case. We define 
the quadratures $X_1=(S_1+S_1^\dagger)/\sqrt{2}$, 
$Y_1=i(S_1-S_1^\dagger)/\sqrt{2}$;
these can be measured e.g. by converting spin excitations to light. 
Correlations between the modes appear due to dynamical evolution and 
squeezing is found in the quadratures of the sum and difference modes $X_\pm=(X_1\pm X_2)/\sqrt{2}$ and $Y_\pm=(Y_1\pm Y_2)\sqrt{2}$. For small number 
of excitations the sum and difference modes obey standard commutation 
relations $\left[X_\alpha,Y_\beta\right]=-i\delta_{\alpha,\beta}$ 
where $\alpha,\beta=+,-\,{\rm or}\,1,2$. A quadrature $Y_\pm$ is squeezed 
when  $\Delta Y_\pm(t)^2<1/2$.  

We find that squeezing is optimal under 
conditions of four-photon resonance 
($\delta_1=\delta_2$) and in the limit of $\eta\gg 1$ (Fig.3).   
Evolution leads to maximum squeezing of $Y_+$ at $t=t^*$ after 
which the growing fluctuations in $X_+$ give rise to increased noise in 
$Y_+$. Note that the number of excitations grows exponentially with 
time (Fig.3c). Specifically, in the case $g_1=g_2$, for $\xi t>1$, we have:
\begin{eqnarray}
\Delta Y_+(t)^2 &=& 1/2\,\left\{e^{-2\xi t} + 2\frac{\gamma_L}{\xi}+
\frac{\kappa/\eta}{\xi}\right. \\ \nonumber
&+& \left.e^{2\xi t}\left(\frac{\gamma_L+\kappa/\eta}{4\xi}\right)^2\right\}
\label{quadt}
\end{eqnarray}
where we have neglected terms of higher order in $\gamma_L/\xi$ and 
$\kappa/\xi$.
The maximum amount of squeezing is obtained after an interaction time 
$t^*$ such that $e^{-2\xi t^*}=(\gamma_L+\kappa/\eta)/4\xi$ and is given by 
$\Delta Y_+^2=(5\gamma_L+3\kappa/\eta)/4\xi$. 
Since both the interaction parameter $\xi$ and the relaxation rate of the
polariton $\gamma_D=\gamma_L+\kappa/\eta$ depend on the single photon
detuning $\Delta$ (Fig.3a), we find that squeezing is optimized for 
the single-photon detuning $\Delta_{opt}=\gamma_{ag}\sqrt{\frac{5|\Omega_1|^2}
{3|\Omega_2|^2}\frac{|g_2|^2N}{\gamma_{ag}\kappa}}$, and  

\begin{equation}
\Delta Y_{+{\rm opt}}^2=\frac{\sqrt{15/4}}{\sqrt{|g_2|^2N/\gamma_{ag}\kappa}}.
\end{equation} 

Note that the denominator is equal to the atomic
density-length product multiplied by an empty cavity finesse and can easily 
exceed $10^4$ even for modest values of the density-length product and
cavity finesse. Note also that the strong coupling regime of cavity QED 
$g\geq{\rm max}[\kappa,\gamma]$ is {\it not} required to achieve strong
correlations, in fact as long as $g^2N\geq 2\kappa\gamma$ squeezing is 
achieved. Furthermore, although a cavity configuration was used for simplicity,
the results of the present analysis remain qualitatively valid in the limit 
of unity finesse, i.e. free space. 
We further emphasize that typical generation rate 
resulting in such optimal squeezing
$\Omega_1\Omega_2/\Delta_{opt}$ can easily be on the order of fraction of
MHz.  In such a case other decoherence mechanisms are negligible.    
Doppler shifts can also be disregarded as long as all fields are 
co-propagating.

To summarize, we have presented a scheme based on the interaction of coherent 
classical light with an optically dense ensemble of atoms that leads to 
effective coherent spin-changing interactions involving pairs of atoms.
We have shown that this process is robust with respect to realistic 
decoherence mechanisms and can result in rapid generation of correlated 
(spin squeezed) atomic ensembles. Furthermore, the resulting spin 
excitations can be easily converted into photons on demand, which
facilitates
applications in quantum information processing. Possible applications 
involving high-precision measurements in atomic clocks can be also
foreseen.
We further note that  extension of this
work into the domain of very large atomic density-length product or
high-finesse
cavities might allow to create maximally spin-squeezed states or 
macroscopic quantum superpositions (``Schrodinger cat'' states). This in
turn  might allow to observe interaction-induced  quantum phase transitions 
\cite{sadchev}. 

We thank M.Fleischhauer, J.I.Cirac, V.Vuletic, S.Yelin and P.Zoller 
for helpful  discussions. This work was supported by the NSF through 
the grant to the 
ITAMP. L.-M.D. acknowledges support from the Austrian and Chinese Science
Foundation.

\def\etal{\textit{et al.}}


\begin{figure}[ht]
\centerline{\epsfig{file=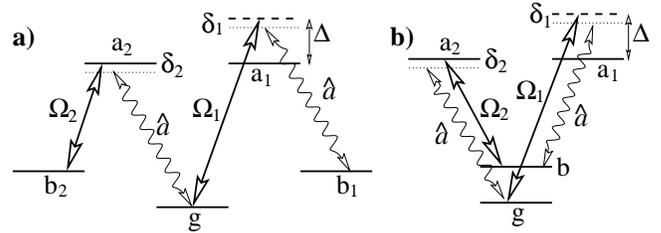,width=8.5cm}}
 \vspace*{2ex}
 \caption{Level scheme for the coherent interaction leading to pairs of atoms 
in (a) different final states $|b_2\rangle$ and $|b_1\rangle$, (b) the same 
final state $|b\rangle$.} 
\end{figure}


\begin{figure}[ht]
\begin{center}
\epsfig{file=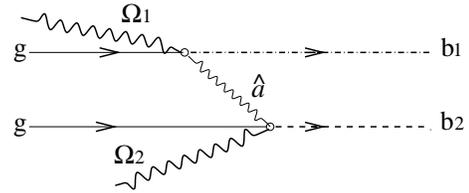,width=6.0cm}
\leavevmode
\end{center}
 \vspace*{2ex}
 \caption{Diagram illustrating coherent atom-atom interaction mediated by 
dark-state polariton, leading to the creation of a pair of spin-flipped atoms.}
\end{figure}


\begin{figure}[ht]
\centerline{\epsfig{file=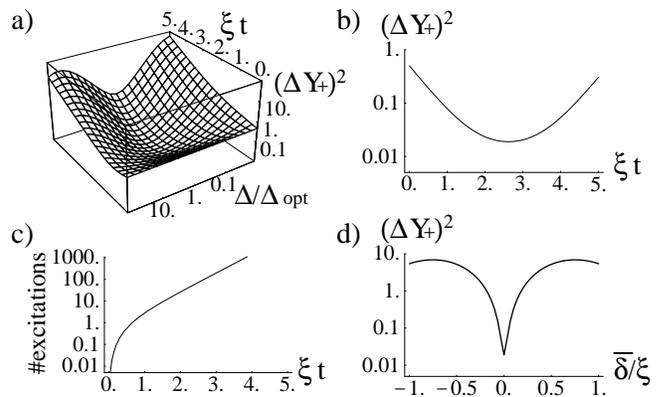,width=8.5cm}}
 \vspace*{2ex}
 \caption{(a) Quadrature variance $\Delta Y_+^2$ vs. single-photon detuning 
$\Delta$ and interaction time $\xi t$, (b) same for $\Delta=\Delta_{opt}$ and 
$\delta_1=\delta_2$ showing maximum squeezing $\Delta Y_+^2\simeq 0.02$ 
(for $\sqrt{g_2^2N/\gamma_{ag}\kappa}=100$), (c) Number of excitations pumped 
in the system vs. time (same conditions as in b) and (d) $\Delta Y_+(t^*)^2$ 
vs. two-photon detuning ${\bar \delta}\equiv (\delta_1-\delta_2)/2$ for 
$\Delta=\Delta_{opt}$ and where $t^*$ gives maximum squeezing.}
\end{figure}


\end{document}